\documentclass[prb,twocolumn,showpacs,preprintnumbers,amsmath,amssymb]{revtex4}
\usepackage{bm} 
\usepackage{graphics}

\begin{document}

\title{Exchange interactions and full magnetization process of multispin nanoclaster Mn$_4$}
\author{A.K. Zvezdin and D.I. Plokhov}
\affiliation{A.M. Prokhorov General Physics Institute of Russian Academy of Sciences, \\ 38 Vavilov Str., 119991, Moscow, Russia}
\date{\today} 

\begin{abstract}
Full magnetization process of magnetic nanocluster Mn$_4$, including all its actual multiplets $S=9/2$, $S=11/2$, $S=13/2$, and $S=15/2$, is theoretically investigated. The formulas needed to determine the exchange constants of cluster Mn$_4$ from experimental data are obtained. It is shown that quantum jumps of magnetization of this nanocluster are in area of megagauss magnetic fields. Remarkable feature of considered nanocluster Mn$_4$ differing it from other multispin molecules is that its exchange spin Hamiltonian supposes exact diagonalization. This provides a reliable basis for comparison of theoretical predictions with experimental data and is important for development of new experimental techniques of high-spin molecule research.
\end{abstract}



\maketitle

\section{\label{sec1} Introduction}

Presently, magnetic molecular nanoclusters draw great attention \cite{ref01}. The clusters are metal-organic molecules containing a number of ions, usually ions of transitive $d$-metals such as Fe, Mn, Co, Ni, etc. or the rare-earth ions, for example Ho, Y, La, etc. They are shortly denoted as X$_n$, where X is Fe, Mn, etc., $n$ is the number of X-ions in the molecule, although this denotation is not complete and unambiguous. The most known and investigated nanoclusters are Fe$_8$, Fe$_{10}$, Fe$_{30}$, Mn$_{12}$, V$_{15}$, etc. Magnetic nanoclusters are also called molecular, or mesoscopic, magnets because they occupy intermediary position between microscopical objects such as atoms and ions, which possess spin moments, and macroscopical magnetic bodies characterized by magnetization, average magnetic moment per volume unit.

Such mesoscopic magnets possess interesting quantum properties \cite{ref02}: a) presence of a quantum hysteresis loop of a single molecule, b) macroscopic quantum tunneling of the full magnetic moment of a molecule, c) effects of a quantum interference at magnetic reversal in a field perpendicular to an easy axis (the effects are related with the Berry-phase) \cite{ref03}. Their properties are rather interesting in the range of submillimeter frequencies \cite{ref04}.

From practical point of view molecular magnets are interesting as building blocks for construction of new materials and nanostructures. They are also considered as perspective objects for quantum computer science \cite{ref05} and magnetic refrigerating tequnics \cite{ref06}.

Among this wide variety of such mesoscopic magnets X$_n$ nanoclusters with formula Mn$_4$O$_3$Br(OAc)$_3$(dbm)$_3$ \cite{ref07}, where OAc is the acetate ion, and dbm is the dybenzoylmethane (see fig. 1) are of special interest. It is known \cite{ref07, ref08, ref09}, that this molecule behaves at low temperatures as an isolated magnet with an easy axis along $C_3$ axis and height of a power barrier of 1.25 meV.

The point symmetry group of molecule Mn$_4$ is (approximately) $C_{3v}$. According to \cite{ref07}, antiferromagnetic interactions of three ions Mn$^{3+}$ ($S=2$) and Mn$^{4+}$ ($S=3/2$), located on axis $C_3$ (passing through Br ions), form ferrimagnetic spin structure with spin of $S=9/2$ in the ground state. The first excited multiplet ($S=7/2$) is separated from the ground one by an energy gap of $\Delta E_1\approx22$ meV.

Magnetic properties of cluster Mn$_4$ in the ground state (multiplet) of $S=9/2$ were investigated in \cite{ref08, ref09}. In the present paper the full magnetization process of Mn$_4$, including all its actual multiplets $S=9/2$, $S=11/2$, $S=13/2$, and $S=15/2$, is theoretically considered, and the formulas needed to determine exchange constants of the cluster from experimental data are obtained. Remarkable and rare feature of considered here nanocluster Mn$_4$, differing it from other multispin molecules, is that its exchange spin Hamiltonian supposes an exact diagonalization. This provides a reliable basis for comparison of theoretical predictions with experimental data and it is important for development of new experimental techniques of research of high-spin molecules.


\section{\label{sec2} Spin Hamiltonian}

A spin Hamiltonian of considered molecule Mn$_4$ within the framework of Heisenberg model reads 
	\begin{equation} 
		H = H_E + H_A, \label{eq01}
	\end{equation}
where exchange Hamiltonian 
	\begin{eqnarray}
		H_E = J_1 (\bm{S}_1 + \bm{S}_2 + \bm{S}_3) \bm{S}_4 + {} \nonumber\\
		{} + J_2 (\bm{S}_1\bm{S}_2 + \bm{S}_3\bm{S}_1 + \bm{S}_2\bm{S}_3) + {} \label{eq02}\\
		{} + \mu _B \bm{B}(g_1(\bm{S}_1 + \bm{S}_2 + \bm{S}_3) + g_2\bm{S}_4) \nonumber.
	\end{eqnarray} 

Here $\bm{S}_1$, $\bm{S}_2$, $\bm{S}_3$, and $\bm{S}_4$ are the spins of ions Mn$^{3+}$ ($S_1=S_2=S_3=2$) and Mn$^{4+}$ ($S_4=3/2$), $g_1$ and $g_2$ are their $g$-factors, $J_1>0$ and $J_2>0$ are exchange constants, and $\bm{B}$ is the external magnetic field.

Term $H_A$ in Eq.~(\ref{eq01}) describes the anisotropy energy that for the ground multiplet ($S=9/2$) can be introduced as \cite{ref08, ref09}
	\begin{eqnarray}
		H_A=D\left[S_z^2-\frac{1}{3}S(S+1)\right] + {} \nonumber\\
		{}+ B_4^0\hat{O}_4^0 + E(S_x^2-S_y^2),
	\end{eqnarray}                                                                                                     
where $D=-62\cdot10^{-6}$ eV, $B_4^0=-6.3\cdot10^{-6}$ eV, and $\left|E\right|=2.1\cdot10^{-6}$ eV \cite{ref08, ref09}. Comparison of these figures with above mentioned value of energy $\Delta E \approx 22$ meV of the first excited multiplet shows that $\left\|H_A\right\|<<\left\|H_E\right\|$. This means that it is possible to be limited to the exchange spin Hamiltonian for calculation of a full magnetization curve. If necessary, the influence of magnetic anisotropy can be taken into account by means of the perturbation theory.


\section{\label{sec3} Addition of spins}

To find the energy spectrum of exchange spin Hamiltonian $H_E$ one can use the following scheme of the spin moments addition:
	\begin{eqnarray}
			\bm{S}_1+\bm{S}_2=\bm{S}_{12}, \nonumber \\
			\bm{S}_{12}+\bm{S}_3=\bm{S}_0, \\
			\bm{S}_0+\bm{S}_4=\bm{S}_t, \nonumber
	\end{eqnarray}
where $\bm{S}_t$ is the total spin of a molecule and $\bm{S}_0$ is the total spin of ions Mn$^{3+}$. The chosen scheme of spin addition is not unique. It is possible to choose other ways, say $\bm{S}_0 = \bm{S}_{13} + \bm{S}_2$ or $\bm{S}_0 = \bm{S}_{23} + \bm{S}_1$. It is known, that such arbitrariness does not influence final results because for any addition scheme the space of the states forms a full set \cite{ref10}. It is necessary to notice that it is usual to use four moment pared addition scheme, for example, $\bm{S}_{12}$ or $\bm{S}_{34}$. Due to natural reasons the above-stated scheme (3 + 1) is more convenient. Transition between various schemes of addition can be made by means of Racah factors or Wigner $9j$-symbols (Fano factors).

The corresponding wave functions being eigenfunctions of operators $S_1^2$, $S_2^2$, $S_{12}^2$, $S_3^2$, $S_0^2$, $S_4^2$, $S_t^2$, and $S_{tz}$ are
	\begin{widetext}
		\begin{equation}
				| S_1 S_2 (S_{12}) S_3 (S_0) S_4 S_t M \rangle = \sum_{m_0m_4} \sum_{m_{12}m_3} \sum_{m_1m_2}
				C^{S_tM}_{S_0m_0S_4m_4} C^{S_0m_0}_{S_{12}m_{12}S_3m_3} C^{S_{12}m_{12}}_{S_1m_1S_2m_2}
				|S_4m_4\rangle |S_1m_1\rangle |S_2m_2\rangle |S_3m_3\rangle, \label{eq5}
		\end{equation}
	\end{widetext}
where $C_{S'm'S''m''}^{Sm}$ are the Clebsch-Gordan coefficients. According to the addition scheme $S_{12}$ takes all integer values from $\left|S_1-S_2\right|$ to $S_1+S_2$, i.e. from 0 to 4. Indexes $m_i$ ($i = $ 0, 1, 2, 3, 4, 12) run all integer (or half-integer for $i = 4$) values from $-S_t$ to $S_t$. The addition rule of $m$-indexes should be taken into account: $m_1+m_2=m_{12}$, etc. Allowable values $S_0$ are shown in Table~\ref{tab1}.

From this table one sees that some values $S_0$ appear in different combinations of $S_{12}$ and $S_3$, i.e. various wave functions correspond to them. This fact is characterized by the factor of degeneration $K(S_0)=2S_0+1$ for $S_0\leq2$ and $K(S_0)=7-S_0$ for $S_0\geq2$. Addition of the moments $\bm{S}_0$ and $\bm{S}_4$ is described by Table~\ref{tab2}. The states with given $S_t$ have degeneration factors $Q(S_t)$ shown in Table~\ref{tab3}.

\begin{table} 
\caption{\label{tab1} Addition of the moments $\bm{S}_{12} + \bm{S}_{3} = \bm{S}_{0}$.}
\begin{ruledtabular}
\begin{tabular}{ccc}
$S_{12}$ & $S_{3}$ & $S_{0}$ \\
\hline
0 & 2 & 2 \\
1 & 2 & 1, 2, 3 \\
2 & 2 & 0, 1, 2, 3, 4 \\
3 & 2 & 1, 2, 3, 4, 5 \\
4 & 2 & 2, 3, 4, 5, 6 \\
\end{tabular}
\end{ruledtabular}
\end{table}

\begin{table} 
\caption{\label{tab2} Addition of the moments $\bm{S}_{0} + \bm{S}_{4} = \bm{S}_{t}$.}
\begin{ruledtabular}
\begin{tabular}{ccc}
$S_{0}$ & $S_{4}$ & $S_{t}$ \\
\hline
0 & $3/2$ & $3/2$ \\
1 & $3/2$ & $1/2$, $3/2$, $5/2$ \\
2 & $3/2$ & $1/2$, $3/2$, $5/2$, $7/2$ \\
3 & $3/2$ & $3/2$, $5/2$, $7/2$, $9/2$ \\
4 & $3/2$ & $5/2$, $7/2$, $9/2$, $11/2$ \\
5 & $3/2$ & $7/2$, $9/2$, $11/2$, $13/2$ \\
6 & $3/2$ & $9/2$, $11/2$, $13/2$, $15/2$ \\
\end{tabular}
\end{ruledtabular}
\end{table}

\begin{table} 
\caption{\label{tab3} Degeneration factors for $S_t$.}
\begin{ruledtabular}
\begin{tabular}{ccccccccc}
$S_{t}$ & $1/2$ & $3/2$ & $5/2$ & $7/2$ & $9/2$ & $11/2$ & $13/2$ & $15/2$ \\
$Q(S_t)$ & 2 & 4 & 4 & 4 & 4 & 3 & 2 & 1 \\
\end{tabular}
\end{ruledtabular}
\end{table}


\section{\label{sec4} Energy spectrum}

Using the considered scheme of spin addition, it is easy to find the energy spectrum of exchange Hamiltonian $H_E$: 
	\begin{eqnarray}
		E(S_t,S_0,M) = \frac{J_1}{2}S_t(S_t+1) - \frac{J_1-J_2}{2}S_0(S_0+1) + {} \nonumber\\
		{} + 2\mu_BMB - \frac{J_1}{2}S_4(S_4+1) - \frac{3}{2}J_2S_1(S_1+1), \label{eq06}
	\end{eqnarray}
where it is supposed that $g_1=g_2=2$. If necessary, distinction between $g_1$ and $g_2$ is easy for taking into account under the perturbation theory. It should be reminded that quantum number $M$ accepts all half-integer values from $-S_t$ to $S_t$. Assuming $B=0$ and normalizing energy of levels on $J_1$ scale one obtains
	\begin{equation}
		\frac{E_0}{J_1} = \epsilon_0(S_t,S_0) =
		\frac{1}{2}S_t(S_t+1) - \frac{\gamma}{2}S_0(S_0+1) \label{eq07}
	\end{equation}
i.e. the spectrum consists of 48 levels completely defined by dimensionless parameter $\gamma=(J_1-J_2)/J_1$.

At this stage it is expedient to address to experimental data. Using Table~\ref{tab2} and Eq.~(\ref{eq07}), it is easy to be convinced that the observable ground state ($S_t=9/2$) can be realized in the given model only if $\gamma>3/4$. Indeed, at increase of $\gamma$ level $\epsilon(S_t=9/2,S_0=6)$ consequently crosses levels $\epsilon(S_t=1/2,S_0=2)$, $\epsilon(S_t=3/2,S_0=3)$, $\epsilon(S_t=5/2,S_0=4)$, and $\epsilon(S_t=7/2,S_0=5)$ at points $\gamma=2/3$, $7/10$, $8/11$, and $3/4$. As it is supposed that $J_1>0$ and $J_2>0$, the area of $\gamma$ change is limited to interval $3/4<\gamma<1$.

Using known position of the first excited multiplet with $S=7/2$ (it is easy to be convinced that it corresponds to $S_0=5$), we receive an additional condition for determining of exchange parameters $J_1$ and $J_2$ (or $J_1$ and $\gamma$):
	\begin{equation}
		\frac{\Delta E}{J_1} =
		\epsilon\left(\frac{7}{2},5\right) - \epsilon\left(\frac{9}{2},6\right) =
		6\gamma-\frac{9}{2}, \label{eq08}
	\end{equation}
$\Delta E = 22$ meV.


\section{\label{sec5} Full magnetization process}

Now we consider a curve of magnetization of molecule Mn$_4$ at $T=0$. We should take into account the above mentioned data on the power spectrum. From Eq.~(\ref{eq06}) follows that the basic condition of system corresponds to $M=-S_t$. It is obvious that there is a consecutive crossing of levels $S_t=9/2\rightarrow11/2$, $11/2\rightarrow13/2$, and $13/2\rightarrow15/2$ with the increase of the magnetic field. According to Table~\ref{tab2}, the value of quantum number $S_0=6$ does not change. The values of magnetic fields at which there are these crossings are easily calculated from Eq.~(\ref{eq06}). They are consequently equal to 
	\begin{eqnarray}
		\frac{9}{2} \rightarrow \frac{11}{2}:\ B_1=\frac{11}{4}\frac{J_1}{\mu_B} \nonumber\\
		\frac{11}{2} \rightarrow \frac{13}{2}:\ B_2=\frac{13}{4}\frac{J_1}{\mu_B} \label{eq09}\\
		\frac{13}{2} \rightarrow \frac{15}{2}:\ B_3=\frac{15}{4}\frac{J_1}{\mu_B} \nonumber 
	\end{eqnarray}

At each point of crossing there is the corresponding change of the ground state accompanied by jump $\Delta M = 2\mu_B$ of the magnetic moment of the molecule. The curve of magnetization is shown in Fig. 2.

Having measured the value of the field corresponding to the first jump of the magnetic moment it is possible to determine value $J_1$ and after that the value of $J_2$ from Eqs. (\ref{eq08}) and (\ref{eq09}).
	\begin{eqnarray}
		J_1 = \frac{4}{11} \mu_B B_1 \\
		J_2 = \frac{1}{4} J_1 - \frac{\Delta E}{6} \nonumber
	\end{eqnarray}

As shown above, parameter $\gamma$ are in interval $(\frac{3}{4},1)$. This enables to estimate the bottom limits for critical fields $B_1$, $B_2$ and $B_3$:

\begin{eqnarray}
	B_1 > \frac{11}{18} \frac{\Delta E}{\mu_B} = 231 \ T \nonumber\\
	B_2 > \frac{13}{18} \frac{\Delta E}{\mu_B} = 273 \ T \\
	B_3 > \frac{ 5}{ 6} \frac{\Delta E}{\mu_B} = 315 \ T \nonumber
\end{eqnarray}

Thus, the full magnetization problem have to be solved by megagauss magnetic field technique \cite{ref11, ref12, ref13}.


\section{\label{sec6} Isotherms of magnetization}

Isotherms of magnetization $M(B,T)$ at any temperatures can be calculated in a standard way by means of the partition function
	\begin{equation}
		Z = \sum_{S_0} K(S_0) \sum_{S_t} Q(S_t) \sum_{M} \exp 
		\left( - \frac{\epsilon_0-bM}{\tau} \right)  	
	\end{equation}
where normalized energy $\epsilon_0(S_t,S_0)$, factors of degeneration $K(S_0)$ and $Q(S_t)$, and parameter $\gamma$ are defined above, $\tau=T/J_1$, $b=2\mu_B/J_1$. The magnetic moment of the system is then determined as \begin{eqnarray} M = \frac{T}{Z} \frac{\partial Z}{\partial B} = g\mu_B \frac{1}{Z} \frac{\partial Z}{\partial b}. \end{eqnarray} Of course, practical calculations of isotherms $M(B,T)$ in actual region of temperatures can be reduced essentially if one uses properties of the power spectrum of a molecule. The corresponding phase diagram is shown in Fig. 3.


\section{\label{sec7} Motivation}

At discussion of motivation of works on molecular magnetism of nanoclusters as one of arguments frequently results the thesis that nanoclusters, being mesoscopic objects, allow to throw light on not quite clear questions of transition from quantum laws, characteristic for atoms and molecules, to classical ones, characteristic for macroscopical bodies. The model of a multispin nanocluster Mn$_4$ investigated in this work is represented rather perspective in this respect since it, being rather simple and admitting an exact diagonalization of its Hamiltonian, is substantially interesting and rich for the description of quantum transformations having well investigated analogues in macroscopical magnetism. For instance, classical analogue of the quantum jumps of the magnetic moment of a molecule considered above (fig. 2) is effect of a turn magnetic sublattices (spin flip) in ferrimagnetics, induced by an external magnetic field \cite{ref14}. The role of sublattices in a case of Mn$_4$ is played by ions Mn$^{3+}$ and Mn$^{4+}$ or their total spins $\bm{S}_0$ and $\bm{S}_4$. The analogue of full magnetization of ferrimagnetic is the magnetic moment of a molecule equal to $2\mu_B\bm{S}_t$.

The Hamiltonian in Eq.~(\ref{eq02}) supposes generalization for any spins $S_i$  ($i$ = 1, 2, 3) and $S_4$. Limiting transition to classics ($S_0 \sim S_4 \sim N \rightarrow \infty $) is naturally performed at the additional assumption $J_1N = const$. The last can be interpreted as the requirement of independence of $N$ the exchange field between sublattices. It is natural to normalize the magnetic moments so that $m_i = \frac{2\mu_BS_i}{N} = const$ were analogues of magnetization sublattices. Then the picture of transition from "ferrimagnetic" spin structure with $S_t = \left| S_0 - S_4 \right|$ to "ferromagnetic" ($S_t = S_0 + S_4$) looks the same as previously described (fig. 222), but the number and amplitude of jumps aspire, accordingly, to $\infty$ and 0 with the increase of $N$, i.e. the transition becomes quasi-continuous. To each jump ($S_t \rightarrow S_t+1$) there corresponds the value of a field at which there is a crossing of $E(S_t)$ and $E(S_t+1)$ levels, i.e. $B_{S_t,S_t+1} = \frac{J_1(S_t+1)}{2\mu_B}$. It also follows that critical fields to which the beginning and the end of spin reorientation correspond are equal
	\begin{equation}
		B_{c1} = \frac{J_1(S_0-S_4+1)}{2\mu_B} =
		\lambda \left| m_0-m_4 \right| + O\left(\frac{1}{N}\right), \label{eq14}
	\end{equation}
	\begin{equation}
		B_{c2} = \frac{J_1(S_0+S_4)}{2\mu_B} = \lambda (m_0+m_4), \label{eq15}
	\end{equation}
where $\lambda = \frac{J_1N}{(2\mu_B)^2}$ is a constant of a molecular field known in the theory of ferrimagnetism.

Eqs. (\ref{eq14}) and (\ref{eq15}) coincide with those from the theory of spin-flip in macroscopical ferrimagnetics \cite{ref14} (the second equation coincides precisely, and the first one coincides to an order of $1/S$). It is natural, that this quantum amendment aspires to the classical formula to 0 with growth of a spin value. But for nanoclusters with small value of a spin in the ground state distinction between the quantum and classical description in this respect is essential. It is possible to tell that this distinction is caused by quantum fluctuations of spins in a vicinity of the critical fields. They are maximal in area of $B_{c1}$ and close to zero in area $B_{c2}$.

Quantum fluctuations are important also for consideration of a question on value of the magnetic moments of spin subsystems (ions Mn$^{3+}$ and Mn$^{4+}$ in this case). In the classical theory it is supposed that the magnetic moments sublattices are constants. Quantum mechanical consideration shows that quantum reduction of sublattices, especially strong near $B_{c1}$, takes place. In work \cite{ref15, ref16} it is shown for a nanocluster Mn$_{12}$ on the basis of the perturbation theory. It is desirable to do this for Mn$_4$, which will be given in the future publications.

Let us note one more interesting difference between spin transformations in "ferrimagnetic" nanocluster and those in macroscopical ferrimagnetics. The last are considered as cooperative phase transitions of second kind \cite{ref14} with corresponding critical anomalies, while the first are essentially quantum transformations occuring in one molecule. Possibly, the role of thermal fluctuations of a parameter in a quantum case is played by quantum spin fluctuations attention was paid on (for a case of nanoclusters such as Mn$_{12}$ and Fe$_8$) in works \cite{ref17, ref18}, which however are executed in a model of giant spin, i.e. in strongly reduced Hilbert space. The model of a multispin system considered in this work with magnetic transformation is interesting and in this respect, as admitting the exact solution in full Hilbert space of the system.

At last, we note one more important question which deserves studying within the framework of the given model. It is known, that energy of chemical binding in molecules depends on a spin state. In this case the magnetic field, changing the spin state, influences also chemical binding of a molecule and thus its configuration. In its turn, change of a configuration influences critical fields of spin transformations. In molecular magnetism the situation can be more various and rich. In this connection we note paper \cite{ref19} in which the effect of downturn of configuration symmetry of a nanocluster in a situation with the spin-degenerated ground state is investigated.


\section{\label{sec8} Acknowledges}

This research was financed by the Russian Foundation for Basic Research (project 08-02-01068).


\section{\label{sec9} Figures}

\includegraphics{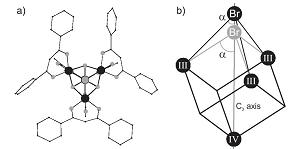}

\textbf{Figure 1.} (a) Molecular structure of the title complex Mn$_4$. View along the approximate C3 axis. For clarity the H atoms are omitted. Mn$^{3+}$ ions are drawn as large black spheres, C and O atoms as small black and grey spheres, respectively. The large grey sphere represents the Br$^{-}$ ion, which obscures the Mn$^{4+}$ ion just behind. (b) Schematic view of the core of Mn$_4$, with III and IV representing Mn$^{3+}$ ($S = 2$) and Mn$^{4+}$ ($S = 3/2$), respectively. The black and grey positions of the Br$^{-}$ ion schematically represent the situation at ambient and high external pressure, respectively.

\includegraphics{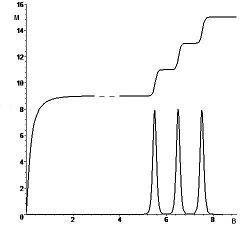}

\textbf{Figure 2.} A curve of full magnetization and quantum steps in nanocluster Mn$_4$.

\includegraphics{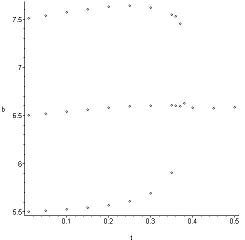}

\textbf{Figure 3.} Phase diagram. For both figures $b = 2\mu_BB/J_1$, $\gamma = 0.8$, $\tau = T/J_1 = 0.05$ (see text for details). 


\end{document}